# MULTIPROCESSOR SYSTEM DEDICATED TO MULTI-ROTOR MINI-UAV CAPABLE OF 3D FLYING

**Adrian-Ioan LIȚĂ*, Ioan PLOTOG*, Lidia DOBRESCU***

*Faculty of Electronics, Telecommunications and Information Technology, POLITEHNICA University of Bucharest, Romania

***Abstract:*** *The paper describes an electronic multiprocessor system that assures functionality of a miniature UAV capable of 3D flying. The apparatus consists of six independently controlled brushless DC motors, each having a propeller attached to it. Since the brushless motor requires complex algorithms in order to achieve maximum torque, efficiency and response time a DSP must be used. All the motors are then controlled by a main microprocessor which is capable of reading sensors (Inertial Measurement Unit (IMU)-orientation and GPS), receiving input commands (remote controller or trajectory plan) and sending independent commands to each of the six motors. The apparatus contains a total of eight microcontrollers: the main unit, the IMU mathematical processor and one microcontroller for each of the six brushless DC motors. Applications for such an apparatus could include not only military, but also search-and-rescue, geodetics, aerial photography and aerial assistance.*
.

## 1. INTRODUCTION

Vertical take-off and landing (VTOL) vehicles have been developed in the past century starting with the helicopter. The main advantages between VTOL vehicles and airplanes is hovering over a small area and being able to operate in areas with no runway. Comparing to a helicopter, having six propellers instead of two highly increases the useful payload, the stability and maneuverability. Multi-rotor systems (with 4, 6 or 8 propellers) can run faster, maintain stability in tougher winds and execute turns much faster (also used in aerobatics) than a traditional helicopter [1]. The main drawback of such systems is that building gas-powered multi-rotor systems is much harder than a classic helicopter.

The apparatus presented by the authors is an electrical powered small-sized vehicle with six vertical propellers.

The applications that can be developed on such a vehicle are of two types: imagery and transportation of small payload. From conception and design point of view, the transportation no needs extra processing. Imagery on the other hand introduces a whole new set of applications, especially when the image can be sent back to base in real time for more complex analysis, according to the mission requests.

## 2. THE MULTIPROCESSOR SYSTEM PROOF OF CONCEPT

**2.1. The multiprocessor system solution.** The authors have chosen brushless DC motors (*BLDC*) as solution for electrical powered of multi-rotor apparatus in order to assure capability to operate with a payload of 4 Kg. The reason for this requirement is that imagery applications demand high-resolution cameras, which are usually heavy.

Taking into account the necessity to implement a complex algorithm capable to achieve maximum torque, efficiency and response time for the motor, a digital signal processor (DSP) type controller was chosen. All the *BLDC* controllers are then controlled by a main microprocessor which is capable of reading sensors (Inertial Measurement Unit (IMU)-orientation and GPS), receiving input commands (remote controller or trajectory plan) and sending independent commands to each of the six motors. The apparatus (Fig. 1) contains a total of eight microcontrollers: the main unit, the IMU mathematical processor and one microcontroller for each of the six *BLDC*.

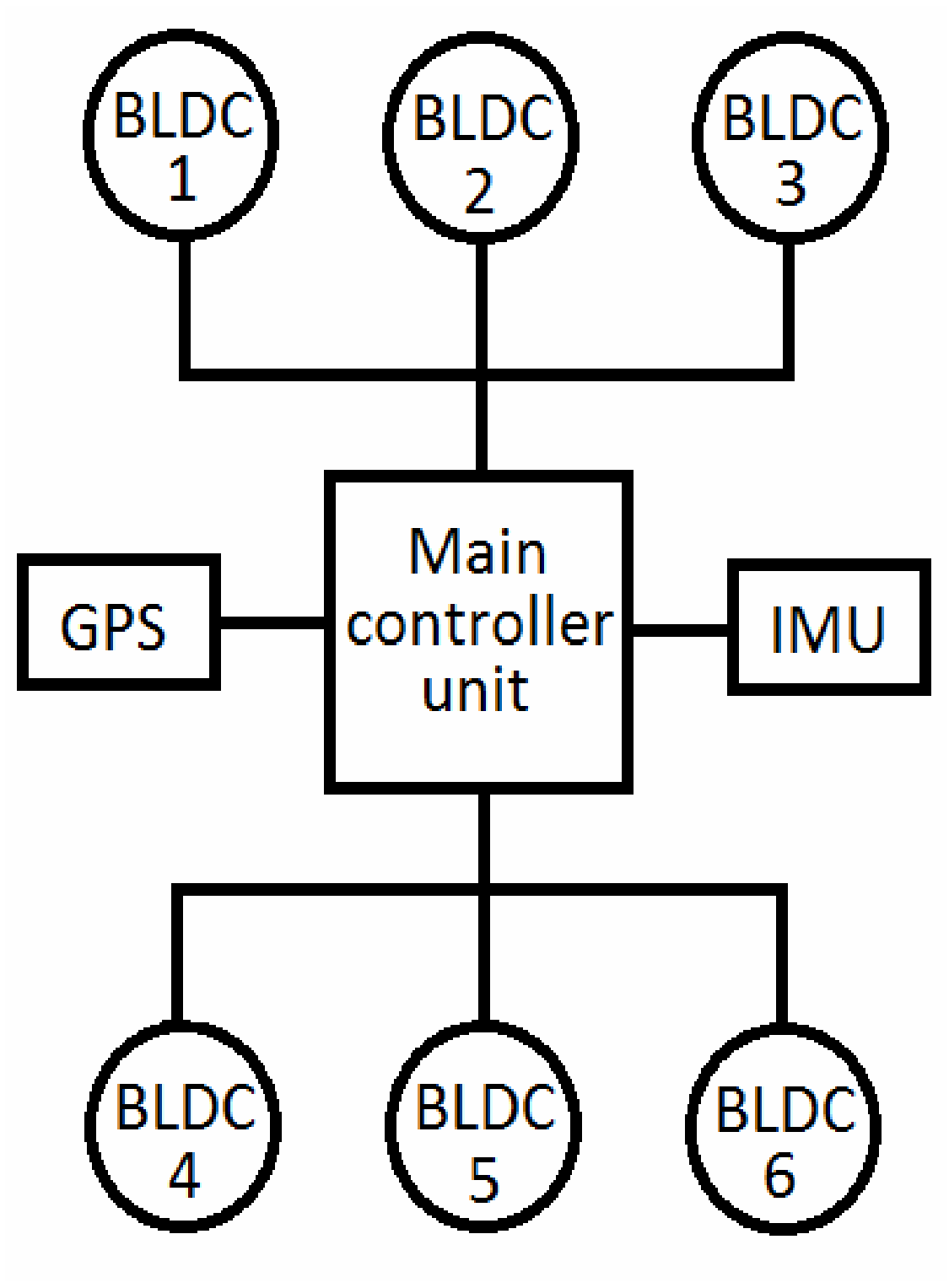

**Fig. 1. Block schematic.**

All the movements are combined and calculated by the main controller unit in order to achieve 3D flying. In (Fig. 2) the scale drawing of the apparatus is presented, noting the fact that the red leg represents the front direction. Also drawn with a circle around each motor is the propeller occupied area. The small arrow on the circle indicates the spinning direction of the propeller.

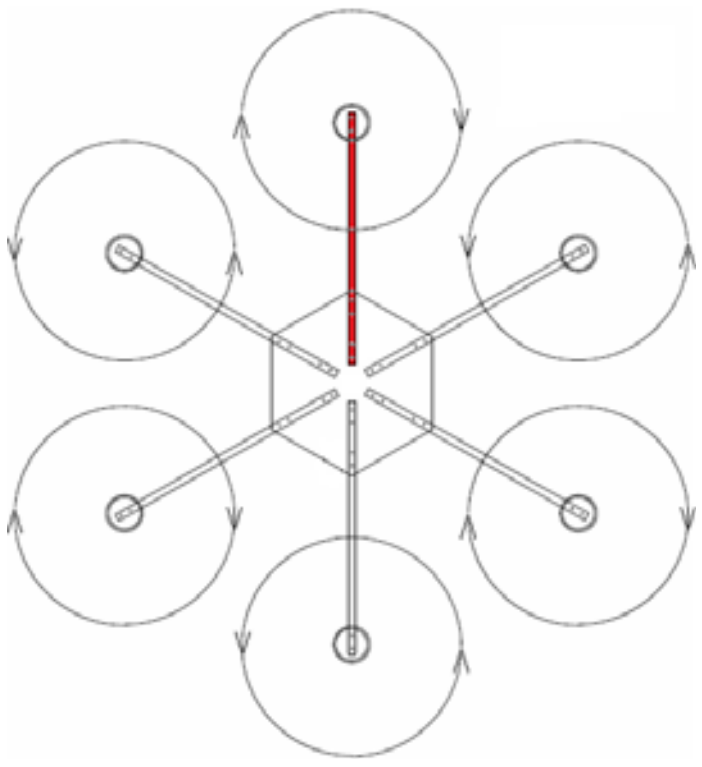

**Fig. 2. Six propellers apparatus general view.**

**2.2. Design of movement control.** While designing the apparatus (next referred as "hexacopter"), in order for it to be able to fully execute three-dimensional maneuvers each movement axis was considered, resulting in three separate moves: roll, pitch and yaw.

For maximum efficiency, all the propellers push air down.

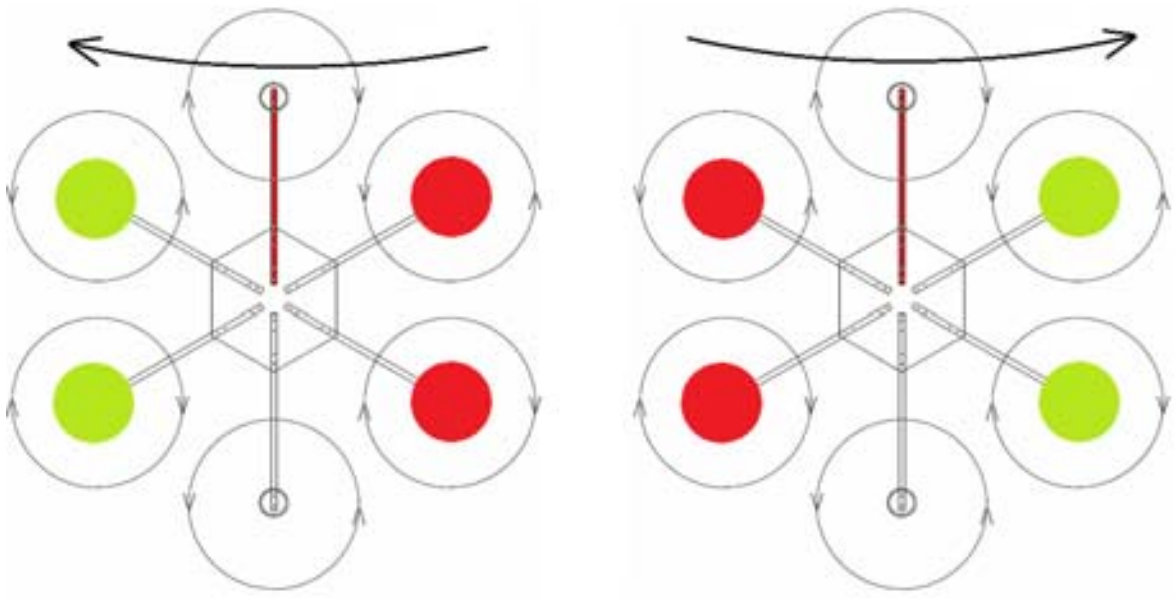

**Fig. 1. Movement on Y axis – Roll.**

The roll and pitch maneuvers are obtainable just by modifying the push force on some of the propellers. In order for the hexacopter to be able to modify the yaw angle and in order for it not to keep spinning at all times due to added momentum, half of the propellers need to spin in one direction and the others in the opposite direction. For best response time, the propellers spin is alternated clockwise and counter clockwise.

Considering the red motor-propeller pair as the front of the hexacopter (parallel to Y Cartesian axis), we can define the roll and pitch movements as following: roll moves the apparatus left and right, while pitch moves it forward or backward. (Fig. 3, 4, 6, 7) depict the propellers spinning speed as follows: transparent – normal spin, red – faster spin, more thrust and green – slower, less thrust. This makes the hexacopter modify its angle and execute the maneuver.

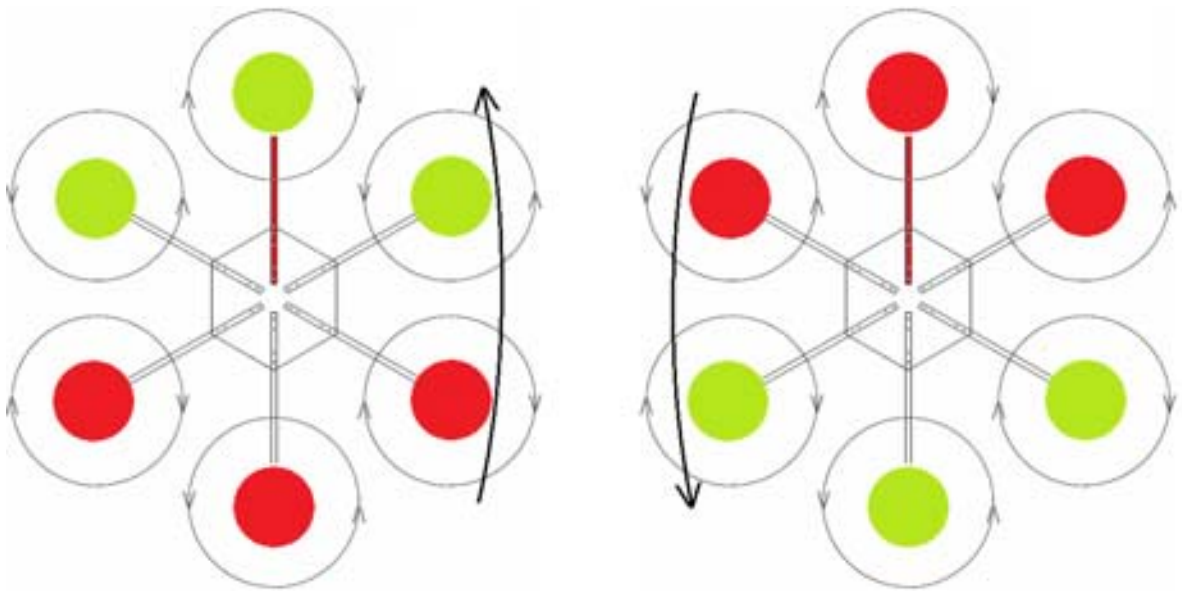

**Fig. 2. Movement on X axis – Pitch.**

It is worth noticing that the roll is done using only 4 out of 6 propellers, while pitch uses all. The reason behind this is that the pitch maneuver dictates the maximum forward speed of the apparatus, which is used more often. (Fig. 5) presents the forward movement of the UAV: red – faster spin, pink – faster spin, but slower than red.

The yaw maneuver is illustrated in (Fig. 6), using all 6 propellers. Fewer propellers may be used for steadier control, with the condition that the total yaw momentum remains constant.

The last described movement is represented by the take-off and landing. Basically all propellers have the same speed, which is either higher (take-off) or lower (landing) than the steady altitude speed. There can be more than one variation, but the fastest is the one illustrated in (Fig. 7).

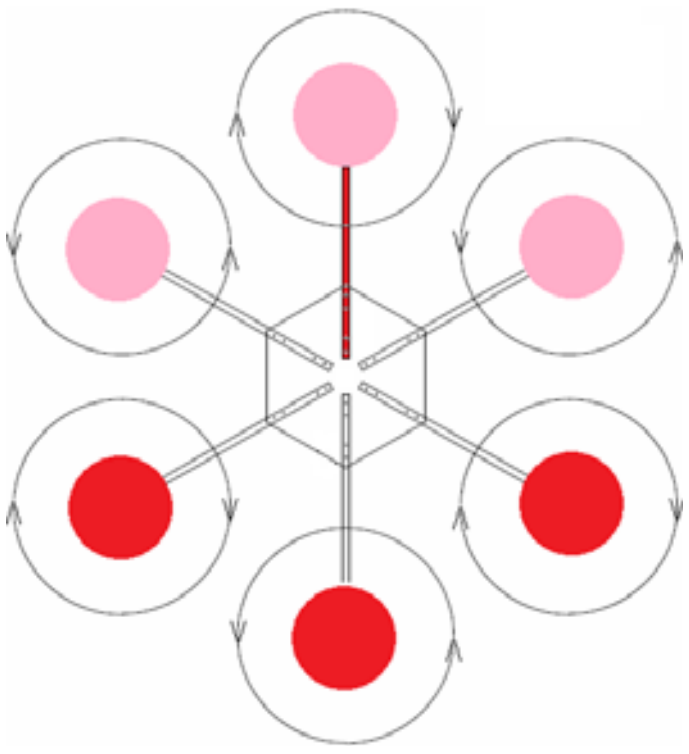

**Fig. 3. Forward Movement.**

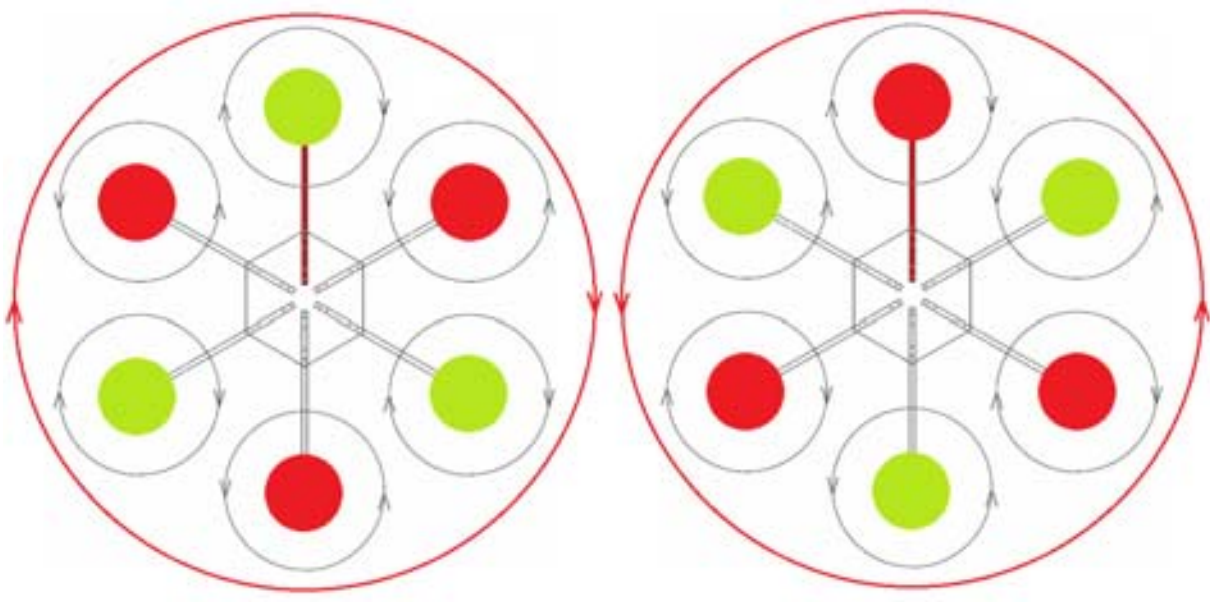

**Fig. 4. Movement on Z axis – Yaw.**

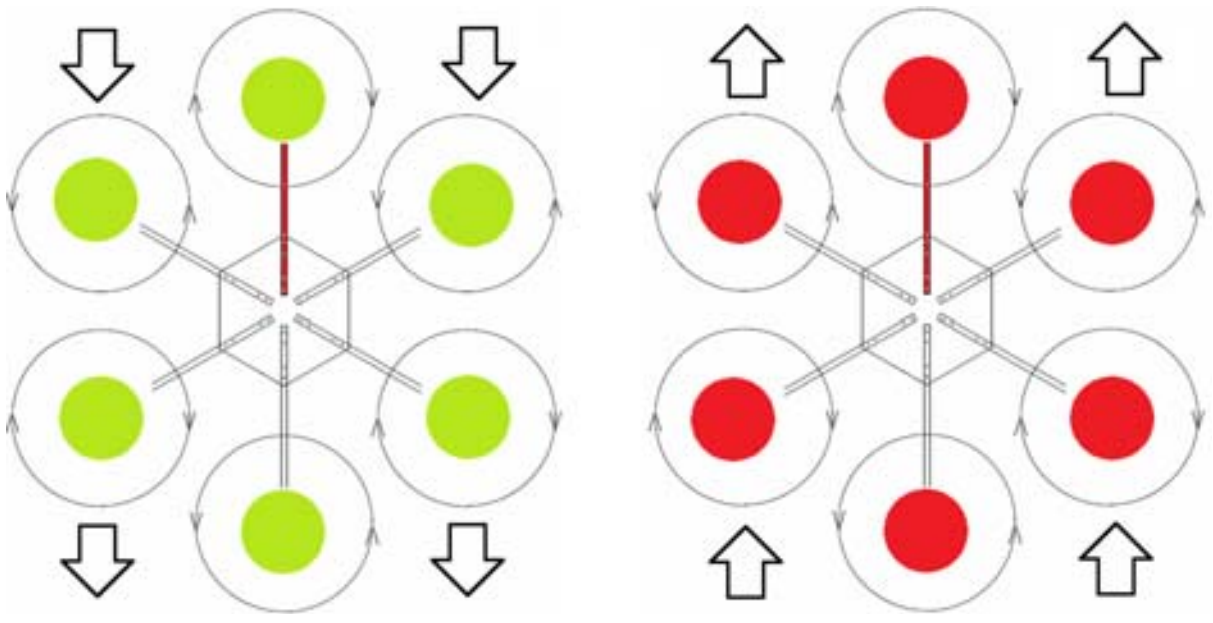

**Fig. 5. Take-off and Landing.**

**2.3 The motor controllers.** The authors have chosen *BLDC* motors NX-4006-530 having speed constant Kv = 530 rpm/v, maxim current = 10A, phase resistance = 75mΩ, phase inductance = 10uH, power = 150W and

weight = 67g [7]. While using brushless motors without any sensor, the usage of a microcontroller with is mandatory. The microcontroller contains a sensorless trapezoidal six-step algorithm which drives six MOS-FET transistors in a three half-bridge configuration used to connect the three phases of the brushless motor.

Six-step trapezoidal commutation (Fig. 8) requires the use of PWM channels. Two half bridges drives two of the phases while the third remains floating and is measured by the ADC module of the microcontroller.

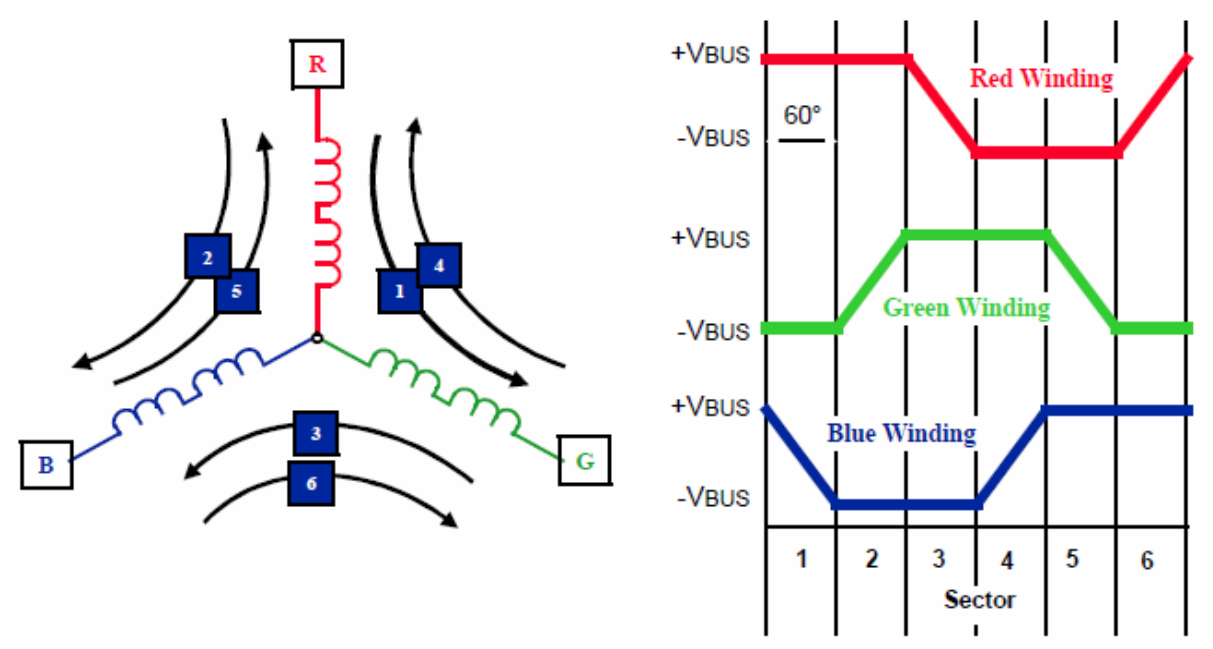

**Fig. 6. Six-step trapezoidal commutation.**

The floating phase is used to compute when the next commutation occurs. This can be usually implemented by two methods, both requiring prior filtering: comparing it either to half the DC voltage or to the reconstructed motor neutral point. The algorithm used in this paper uses the first method as it requires fewer electronic components and makes full use of the microcontroller's DSP functions.

Comparing the floating phase voltage (also known as back electro-magnetic force – *BEMF*) to half the DC voltage will result in finding the zero-crossing point (Fig. 9) of the voltage. The zero-crossing point is the exact moment at which the floating phase voltage is equal to DC bus voltage divided by two, and is the exact moment at which the rotor moved 30° out of 60° of a sector. By measuring the time between the sector change and zero crossing moment (30°), the algorithm then waits this exact amount of time before changing the sector again. This approach is used to replace the sensor by using the floating phase as a sensor. By using PWM modulation to control the MOS-FET transistors, the motor speed is controlled by increasing or decreasing the duty-cycle. In order for the commutation not to create acoustical disturbance, the PWM frequency used is 20 KHz. High frequency require more expensive MOS-FET transistors to keep the losses to minimum. The duty-cycle is set by a PI (proportional integral) controller, which receives the speed as the measured parameter and the desired speed from the main controller unit as the input parameter.

The communication between the main controller unit and the motor controllers is done via I2C protocol.

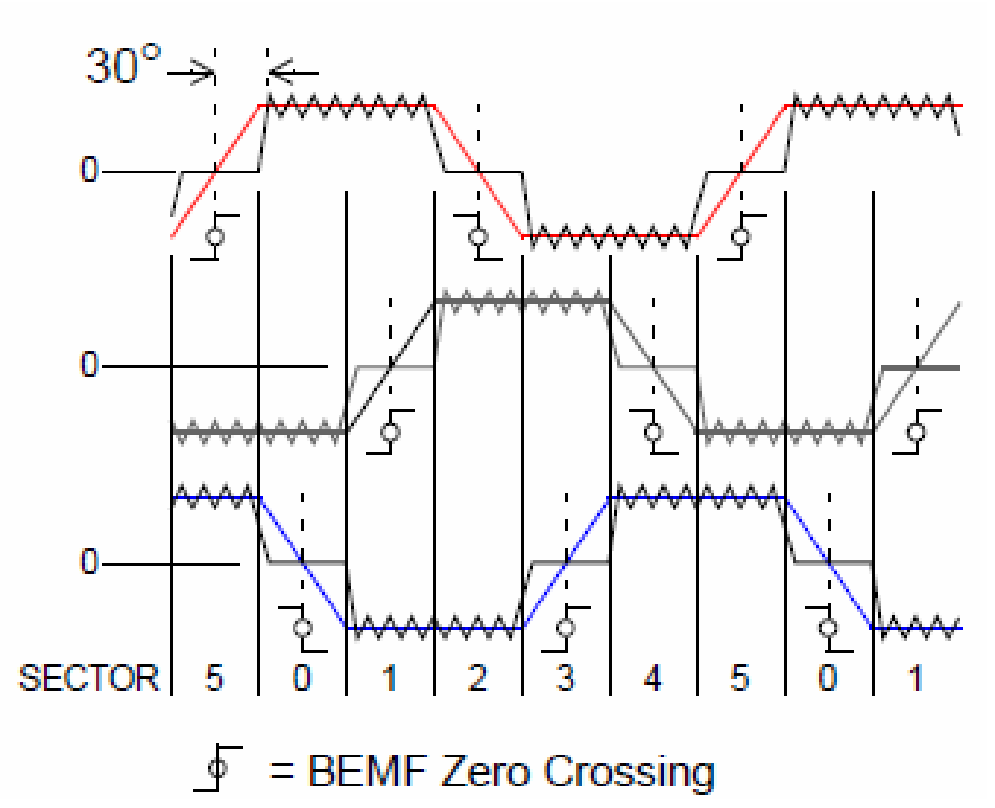

**Fig. 7. Zero-Crossing detection.**

The microcontroller used in this case is the dsPIC33EP64MC502 made by Microchip, for its motor control dedicated PWM channels, its 10-bit ADC module and its communication protocols (I2C and also CAN, leaving room for further implementation) [2]. The devices consume less than 100 mW of power, which minimizes the losses.

**2.4 The main controller unit.** The main controller unit is responsible for sensor data acquisition, wireless communications with the base station, communication with the motor controllers and for the main movement computations. It contains a 32 bit microcontroller capable of making all the calculations required in order for the apparatus to be stable and to have a fast response. The main controller unit also has access to the GPS sensor. The main controller unit has the possibility of sending all its data to another system (faster, such as a PC) via wireless link or directly via UART or SPI. This was designed to allow running of higher complexity applications such as automatic search and rescue, or monitoring. They require a color camera and image analysis software, which is very demanding and can't be ran on the main controller unit. The interface allows full control of the hexacopter by the 3rd party computer, but still allowing for manual override of the commands if needed. The processor used for the main controller unit is the PIC32MX795F512L made by Microchip. It offers a lot of communication protocols, 4 UART modules, 4 I2C modules, and over 100 MIPS of processing power on 32 bits mathematical sets, while requiring less than 400 mW of power [2].

The main sensor is the inertial measurement unit (IMU) which uses a three-axis accelerometer, a three-axis gyroscope and a three-axis magnetometer in order to compute the angles of roll, pitch and yaw. Initially, when the hexacopter is not started, all angles are reset in order to set the forward direction. The sensor uses UART communication to send the data back to the main controller unit.

## 3. THE EXPERIMENTAL RESULTS AND DISCUSSIONS

**3.1. Build.** As seen in (Fig. 10), the six motor-propeller pairs are connected to the main unit through carbon-fiber tubes to a central aluminum hexagonal plate. Carbon-fiber and aluminum were chosen for being very lightweight materials with increased durability and strength. The total weight of the apparatus is 1.8 Kg. The central plate was chosen to be hexagonal instead of round in order to provide higher contact to carbon-fiber tubes. Also, for increased strength, the carbon-fiber tube profile is square instead of round. The tube houses the wiring needed to provide power and commands to the motor controllers.

**3.2. Lift and power consumption tests.** Tests have shown that each motor-propeller pair is capable of pushing air up to a thrust capable of lifting 1 Kg / pair. Having 6 pairs and a 1.8 Kg apparatus gives us a total of about 4 Kg of useful payload, exactly as designed.

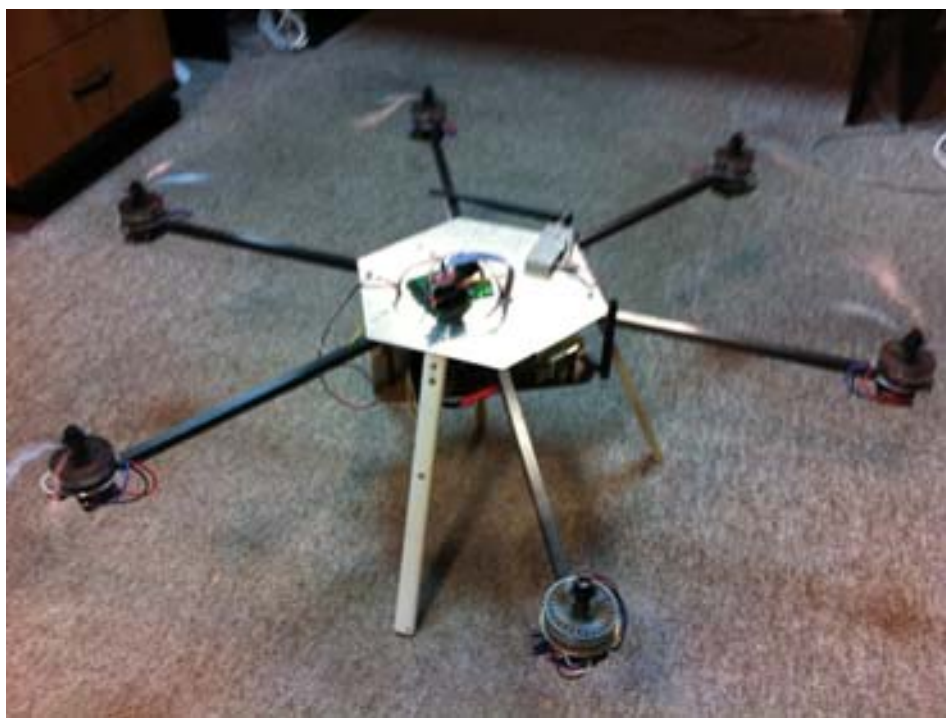

**Fig. 8. Hexacopter.**

The brushless DC motor controllers provides very fast response, low commutation noise and excellent performance (Fig. 11). New speed information is fed to the motor controller with a frequency of 500 Hz.

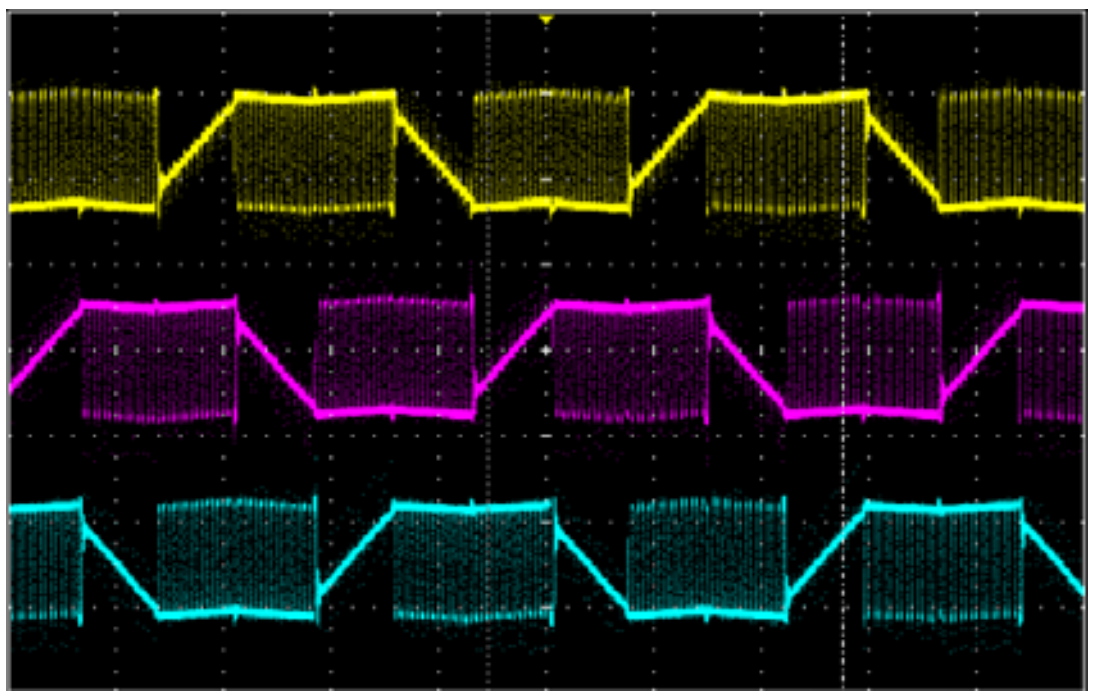

**Fig. 9. Motor BEMF and driving voltage.**

The total power consumption of the hexacopter while standing, with no motors running was of 850 mW, which, with a battery of 16.8V and 5800 mAh (a total of 97.4 Wh) is less than 1%.

Each of the motors running in full thrust (when able to lift approximately 1 Kg each) had a current consumption of 9.8 A at 16.8V. With the hexacopter running in full thrust, this would give about 6 minutes of flight time. However, while hovering (keeping a steady altitude in no-wind conditions) with no payload, the total power consumption is 170W, giving a total flight time of over 30 minutes. The exact case is somewhere between 6 and 30 minutes, depending on the payload: when the apparatus is required to have extra payload (1 Kg), it is still fast in response, but the total consumption is 290W, which means about 18-20 minutes of flight. When maximum load is applied, the response time is slow, since there isn't a lot of reserve power to make the fast movements: all the power is consumed to lift and keep the payload.

The BLDC controller (Fig. 12) was designed in respect of Design for Manufacturing concept [8], so it would fit below the motor, in order to save space and increase efficiency.

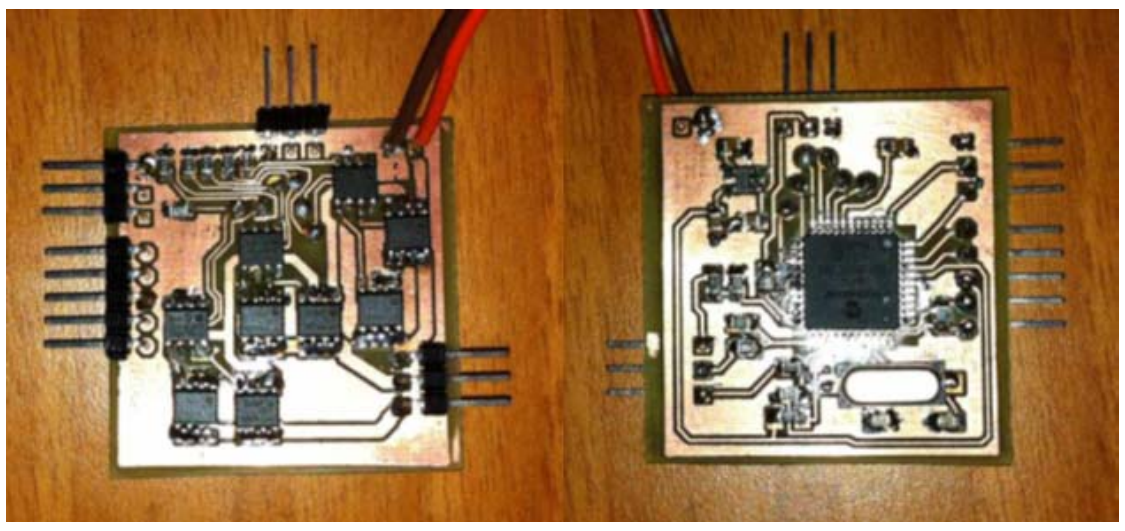

**Fig. 10. BLDC controller board – all layers**

Further development require designing new algorithms which can be applied to the hexacopter in order to automatically perform complex tasks, most in the field of digital image processing: automatic search for certain objects on an area, monitoring perimeter, etc.

## 4. CONCLUSIONS & ACKNOWLEDGMENT

**4.1. Conclusions.** The main controller unit initially has a simple PID controller to feed the output for the motor controllers. Though the response time is good, some oscillations are present. Tuning the PID loop parameters is mandatory, and the PID was added complexity by adding an auto-tuning sequence to be executed on every start of the hexacopter.

The ability to execute 3D maneuvers gives virtually any possibility to any application which requires 3D flight. More complex algorithms can be implemented when using more than one hexacopter to perform a collaborative work. Other example would be that the estimated flight time is around 20 minutes, and the charge time is around 1.5-2 hours, meaning that in order to fully monitor a perimeter, at least four are needed, one to be operational and the others to charge.

**4.2. Acknowledgment**. This work has partial supported by the SIOPTEF project (PN-II-PT-PCCA-2011-3, C-121/2012).

## REFERENCES

1. Gablehouse C. Helicopters and autogiros: a chronicle of rotating-wing aircraft Lippincott (1967)
2. Lita, A., Cheles M. *AN1160 Sensorless BLDC Control with Back-EMF Filtering Using a Majority Function* [online]. Available: http://www.microchip.com
3. Alamio A., Artale V., et all. PID Controller Applied to Hexacopter Flight *Journal of Intelligent & Robotic Systems.* Issue 1-4 (Jan 2014).
4. Oursland J., The design and implementation of a quadrotor flight controller using the QUEST algorithm *South Dakota School of Mines and Technology* (2010)
5. Markley, F. L., Mortari D. Quaternion altitude estimation using vector observations *The Journal of the Astronautical Sciences* (2000)
6. Erginer B., Altug E. Modeling and PD control of a quadrotor VTOL vehicle *IEEE Intelligent Vehicles Symposium* (2007)
7. NX-4006-530kv Brushless Quadcopter Motor. Source. [online]. Available: http://www.hobbyking.com/hobbyking/store/__17923__NX_4006_530kv_Brushless_Quadcopter_Motor.html.
8. Plotog I., Marin A., Boanta L., Model of assembling process for electronic parts integrated in mechatronic products, The Romanian Review Precision Mechanics, Optics & Mechatronics, 2013, No. 44, ISSN 1584 – 5982.